# Ultraconfined THz Phonon Polaritons in Hafnium Dichalcogenides

R. A. Kowalski[1§], N. S. Mueller[2§], G. Álvarez-Pérez[2,3,4], M. Obst[5,6], K. Diaz-Granados[1], G. Carini[2], A. Senarath[1], S. Dixit[1], R. Niemann[2], R. B. Iyer[7], F. G. Kaps[5,6], J. Wetzel[5,6], J. M. Klopf[8], I. I. Kravchenko[9], M. Wolf[2], T. G. Folland[7], L. M. Eng[5,6], S. C. Kehr[5,6], P. Alonso-Gonzalez[3], A. Paarmann[2], J.D. Caldwell[1]

[1]Vanderbilt University, Nashville, TN, USA, [2]Fritz Haber Institute, Berlin, Germany, [3]Department of Physics, University of Oviedo, Oviedo, Spain, [4]Istituto Italiano di Tecnologia, Arnesano, Italy, [5]Institute of Applied Physics, TUD Dresden University of Technology, Dresden, Germany, [6]ct.qmat - Excellence Cluster TU Dresden-Würzburg, [7]University of Iowa, Iowa City, IO, USA [8]Helmholtz-Zentrum Dresden Rossendorf, Germany, [9]Oak Ridge National Laboratory, Oak Ridge, TN, USA

§Indicates equal contributions

**Abstract**

The confinement of electromagnetic radiation to subwavelength scales relies on strong light-matter interactions. In the infrared (IR) and terahertz (THz) spectral ranges, phonon polaritons are commonly employed to achieve extremely subdiffractional light confinement, with much lower losses as compared to plasmon polaritons. Among these, hyperbolic phonon polaritons in anisotropic materials offer a highly promising platform for light confinement, which, however, typically plateaus at values of $\lambda_0/100$, with $\lambda_0$ being the free-space incident wavelength. In this study, we report on ultraconfined phonon polaritons in hafnium-based dichalcogenides with confinement factors exceeding $\lambda_0/250$ in the terahertz spectral range. This extreme light compression within deeply sub-wavelength thin films is enabled by the unprecedented magnitude of the light-matter coupling strength in these compounds, and the natural hyperbolicity of $HfSe_2$ in particular. Our findings emphasize the critical role of light-matter coupling for polariton confinement, which for phonon polaritons in polar dielectrics is dictated by the transverse-longitudinal optic phonon energy splitting. Our results demonstrate transition metal dichalcogenides as an enabling platform for THz nanophotonic applications that push the limits of light control.

**Main**

A significant focus of the field of nanophotonics is the ability to confine electromagnetic energy to highly subdiffractional length scales. This enables increased electric field intensities that can cause enhanced light-matter interactions, such as surface-enhanced Raman scattering,[1] optical nonlinearities,[2,3] and infrared absorption,[4] among others. Significant light compression has been achieved by reducing the light-matter interaction volume using resonant cavities[5] and atomic-layer thicknesses of materials,[6–9]; nonetheless, this increased optical confinement is severely hampered in many cases due to high losses.[10] Polaritons, hybrid light-matter quasiparticles, have facilitated extreme light confinement[11–13] enabling subdiffractional imaging,[14–16] nanoscale spectroscopy,[4,17] and nanophotonic circuits.[18] At IR wavelengths, optically active polar lattice vibrations couple with light creating phonon polaritons (PhPs) that can propagate with high momenta ($k$),[19] spectral tunability,[20] and propagation directionality[21] with the advantage of significantly reduced losses with respect to plasmon polaritons in conductors.[22,23]

Surface PhPs can be supported within the Reststrahlen band (RB) of polar dielectric materials, which is defined as the frequency range between the transverse optic (TO) and longitudinal optic (LO) phonons, within which the real part of the dielectric permittivity tensor is negative ($\varepsilon < 0$).[24] The modes are comprised of evanescent waves bound to the surface of the polaritonic medium with subdiffractional polariton wavelengths ($\lambda_p$).[25,26] Moving from bulk to thin films, the surface-bound waves hybridize and confine further (~$\lambda_0/30$)[27,28] as the thickness decreases ($\propto 1/d$).[29,30] Along this line, efforts to achieve maximum confinement ($\lambda_0/\lambda_p$) have been focused on using atomically-thin films[8,30] or identifying anisotropic materials that exhibit an extreme form of birefringence called hyperbolicity,[31–33] which occurs in materials with dielectric permittivities of opposite sign along different crystallographic directions. In hexagonal crystals, such as in hexagonal boron nitride (hBN), hyperbolicity is realized along in- and out-of-plane crystallographic directions resulting in $\varepsilon_\parallel \cdot \varepsilon_\perp < 0$. Hyperbolicity is commonly found in van der Waals (vdW) crystals, which are comprised of covalently-bonded two-dimensional (2D) sheets along the tangential plane ($xy$, labeled as $t$) that are stacked axially ($z$) and held together by weaker vdW bonds.[11–13,16] Phonons in uniaxial vdW crystals with polarizations oriented in the tangential and axial directions are therefore nondegenerate and induce a natural anisotropy resulting in nonidentical permittivities ($\varepsilon_t \neq \varepsilon_z$). The large confinement occurs due to the extraordinary modes, or hyperbolic PhPs that possess nominally unrestricted magnitudes of $k$ that traverse through the volume of the material.[33] Furthermore, in ultrathin hyperbolic films, where the thickness is far below the free-space wavelength ($d << \lambda_0$), the PhP confinement scales as $\propto 1/d$ down to the monolayer limit.[30] Despite reports of exceptionally high confinement ($\lambda_0/500$),[8] the losses encountered at these atomic scales far outweigh the intensity required for the PhP to propagate sufficiently for many applications. At more modest film thicknesses ($d$ = 10-100 nm), PhP propagation is achievable but not without sacrificing confinement to values approximately $\lambda_0/100$.[32,33]

Phonon polaritonic materials with high confinement, such as alpha-phase molybdenum trioxide (α-MoO$_3$)[21,32,34,35] and hBN,[13,33] have large LO-TO phonon splitting (> 100 cm$^{-1}$).[33,36] Broad RBs play an important role in confinement as the large bandwidth allows the PhP to extend to high $\boldsymbol{k}$ before dissipating and with non-zero group velocity ($v_g = \frac{d\omega}{d\boldsymbol{k}}$), with respect to other phonon polariton materials with narrower RBs. Furthermore, the magnitude of the light-matter coupling strength is directly related to LO-TO phonon energy splitting which is a function of the Born effective charge.[22] However, this is also an absolute quantity that tends to be larger for mid-infrared (mid-IR) phonons compared to those resonant within the THz. A polar dielectric material with a similar RB width to hBN at THz frequencies would therefore exhibit drastically increased light-matter coupling strengths and stronger optical confinement than other THz resonant materials.[31,34,37–39] Exceptionally large Born effective charges of hafnium-based dichalcogenides (HfDCs) give rise to their large LO-TO energy splittings,[40,41] while the heavy Hf-ion mass places their natural optic phonon frequencies in the THz range. The natural anisotropy of these 2D vdW crystals, accompanied with their strong light-matter coupling strength, highlights them as excellent candidates for exploration of ultraconfined PhPs in the THz spectral regime.[40]

In this work, we observe ultrahigh confinement of THz light within the HfDCs, hafnium disulfide (HfS$_2$) and hafnium diselenide (HfSe$_2$), with PhP wavelengths, $\lambda_p$, significantly compressed with respect to the free-space wavelength ($\lambda_0/\lambda_p$ > 250). We examine the THz near-field response of this unique set of materials using a scattering-type scanning near-field optical microscope (s-SNOM) with the light provided by a free-electron laser (FEL). While they share hexagonal crystal structures, HfS$_2$ and HfSe$_2$ have different infrared dielectric permittivities[40] that separate the predominant type of PhPs that can be supported into elliptic and hyperbolic, respectively. In this regard, elliptic refers to an anisotropic response whereby both in- and out-of-plane dielectric permittivities are different in magnitude, but both being negative, whereas for hyperbolicity one of these directions is negative, while the other is positive.[13] Interestingly, we observe very strong confinement of PhPs in ultrathin films of both materials for extremely small film thickness $d$ as compared to the long THz wavelengths ($\lambda_0/d \sim 10^3$). This is possible since the LO-TO splitting of HfS$_2$ and HfSe$_2$ is on par with high-confinement mid-IR phonon polaritonic materials. However, when normalizing by the resonant frequency ($\omega_{TO}$), the HfDCs display significantly larger light-matter coupling strengths, which are responsible for the observed ultrahigh-$k$ PhPs. Furthermore, we demonstrate THz hyperlensing by placing a metal nanoantenna below a hyperbolic HfSe$_2$ flake, enabling the observation of hyperbolic rays that can be employed for deeply sub-diffractional THz imaging. The observation of ultraconfined THz polaritons in Hf-based vdW crystals demonstrates exceptionally high promise for deeply subdiffractional THz nanophotonic components.

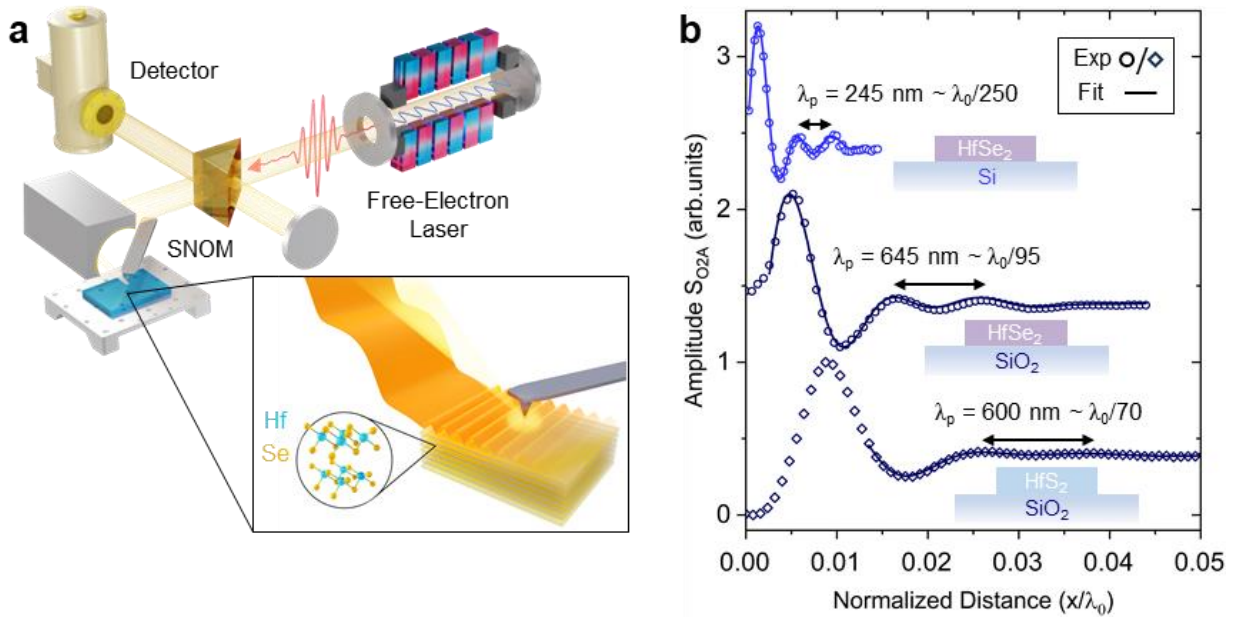

**Figure 1:** (a) Schematic of the experimental s-SNOM apparatus. A tunable FEL generates narrowband THz radiation that is focused onto the s-SNOM tip and is back-scattered into a photoconducting detector. Inset: THz light scatters off the edge of a HfDC flake, launching PhPs, which are then coupled out to free-space by the s-SNOM tip. Additionally, the layered crystal structure of vdW-bonded HfSe$_2$ is shown. (b) Line profiles, normalized by the free-space wavelength $\lambda_0$, of the raw s-SNOM amplitude (S$_{O2A}$, offset for clarity) are plotted showing the propagation of edge-launched PhPs (with edge to the left) for three samples: HfSe$_2$ on Si ($d$ = 47 nm, $\lambda_0$ = 61.7 µm, blue circles), HfSe$_2$ on SiO$_2$ ($d$ = 68 nm, $\lambda_0$ = 60.7 µm, dark blue circles), and HfS$_2$ on SiO$_2$ ($d$ = 68 nm, $\lambda_0$ = 41.7 µm, dark blue diamonds). Each profile was fitted (SI Section S2) to extract the polariton wavelengths ($\lambda_p$), which are labeled along with the corresponding confinement factors.

Directly observing highly confined PhPs in the far-field might be challenging due to the intrinsic momentum mismatch with free-space light, which is orders of magnitude different from the THz and the nanoscale polariton wavelengths supported. Near-field microscopy using s-SNOM is ideally suited to access and directly image highly confined PhP modes.[42–44] Notably, the accessible momentum range in this work is experimentally limited, rather than the material response, yet still the nanotip size employed in s-SNOM provides access to even larger confinement factors in the THz than possible in the mid-IR. We use a tunable FEL to generate narrowband THz radiation[45–47] that when scattered off the s-SNOM tip and/or the edges of the HfDC flakes results in scattered fields with momenta sufficient to launch and detect highly confined PhP modes. An image of the PhP is obtained by rastering the tip across the surface of the flake and collecting the scattered light at each point. A schematic of the s-SNOM apparatus and of PhPs launching from the edge of a flake are shown in Figure 1a.

One-dimensional line profiles are extracted from the s-SNOM images (SI Section S1) perpendicular to the edge of the flakes and plotted in Figure 1b for three sample configurations: HfS$_2$ on a SiO$_2$ substrate (dark blue diamonds), HfSe$_2$/SiO$_2$ (dark blue circles), and HfSe$_2$/Si (blue circles), where each line profile is normalized on the $x$-axis by $\lambda_0$. Fits of the line profiles show that the edge-launched polaritons dominate the response that are then scattered to the far-field by the s-SNOM tip (SI Section S2). The peak-to-peak distance between fringes therefore corresponds to the polariton wavelength.[36,48] Deeply subdiffractional modes are observed in HfS$_2$/SiO$_2$ ($\lambda_p$ = 600 nm, $\lambda_0$ = 41.7 μm, $d$ = 68 nm) and HfSe$_2$/SiO$_2$ ($\lambda_p$ = 645 nm, $\lambda_0$ = 60.7 μm, $d$ = 68 nm) where the PhP confinement factors ($\lambda_0/\lambda_p$) are 70 and 95, respectively, as shown in Figure 1b. Notably, we observe an even larger confinement of >250 for HfSe$_2$/Si with a polariton wavelength of 245 nm ($\lambda_0$ = 61.7 μm, d = 47 nm) (blue curve Figure 1b). Such extreme PhP confinement, without enhancement from nano-resonators, has previously only been reported in mono- and few-layer materials with high propagation losses.[8] However, the thicknesses ($d$) of the HfDC flakes explored here are on the order of tens of nanometers, thus preserving long polariton propagation.

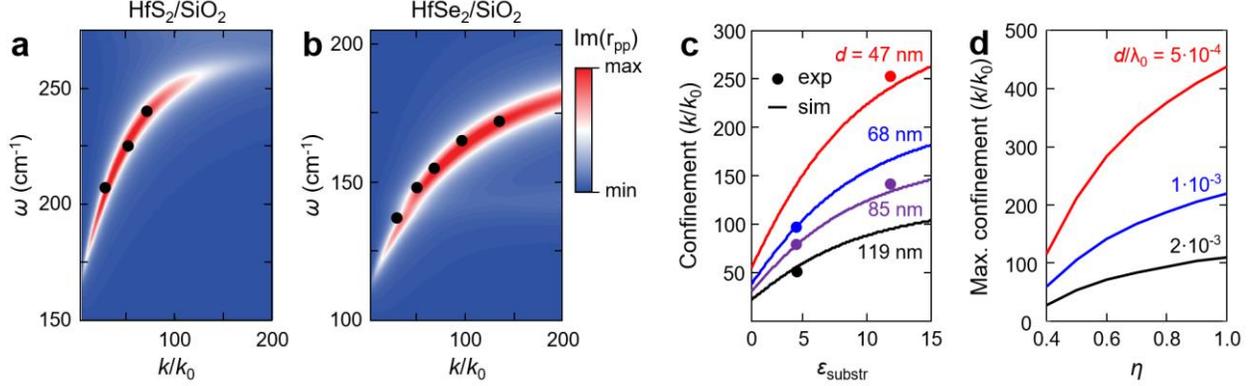

**Figure 2:** The PhP dispersions of (a) $HfS_2/SiO_2$ ($d$ = 68 nm) and (b) $HfSe_2/SiO_2$ ($d$ = 68 nm) thin films. Experimental data (black circles) are plotted over the simulated $Im(r_{pp})$ illustrating the extremely large $k$ of the propagating PhPs. (c) Light confinement of the PhPs in $HfSe_2$ as a function of substrate permittivity ($\varepsilon_{sub}$). Confinement factors are calculated using the TMM with flake thicknesses ($d$ = 47, 68, 85, and 119 nm) used in experimental s-SNOM imaging. Experimental values (circles) from $HfSe_2/SiO_2$ ($\varepsilon_{sub}$ = 4, $\omega_0$ = 165 cm$^{-1}$) and $HfSe_2/Si$ ($\varepsilon_{sub}$ = 12, $\omega_0$ = 162 cm$^{-1}$) are plotted with the simulated values (solid lines, $\omega_0$ = 163.5 cm$^{-1}$) for different $HfSe_2$ thicknesses (colors, see legend). (d) Maximum confinement of thin film polaritons (circles) of a type-II hyperbolic model material as a function of normalized light-matter coupling strength ($\eta$) at three fixed thickness to free-space wavelength ratios ($d/\lambda_0$ = 2·10$^{-3}$, black; 1·10$^{-3}$, blue; and 5·10$^{-4}$, red); see SI Section S5 for details.

Ultrahigh-$k$ PhPs are typically only observed in hyperbolic materials, and the ultraconfined modes observed in $HfSe_2$ are indeed hyperbolic. However, we also observe high-$k$ propagating PhPs in $HfS_2$ ($\lambda_0/70$), which is elliptic in this spectral range.[40] To further understand such large confinement of both elliptic and hyperbolic PhPs, we examine the differences between polariton dispersions of $HfS_2$ and $HfSe_2$ based on experiments and simulations. For this purpose, we collect s-SNOM images at several excitation frequencies ($\omega_0 = ck_0$) and plot the energy-momentum dispersions where $k/k_0 = \lambda_0/\lambda_p$ (Figure 2a,b). The experimental dispersions (black circles) are in good agreement with the simulated imaginary component of the *p*-polarized reflectivity $Im(r_{pp})$, calculated using the transfer matrix method (TMM)[49] based on the dielectric permittivities extracted from far-field experiments.[40] A dispersive mode is visible in both $HfS_2/SiO_2$ (Figure 2a) and $HfSe_2/SiO_2$ (Figure 2b), with the dispersions extending between 160-275 cm$^{-1}$ and 120-190 cm$^{-1}$, respectively. The two PhPs start dispersing similarly at lower frequencies before curving toward larger values of $k$. There, however, the hyperbolic mode maintains a finite slope as well as its $Im(r_{pp})$ intensity, in contrast to its elliptic counterpart that decays in intensity due to higher losses and becomes non-dispersive. Considering the long THz wavelengths and the nanoscale sample thicknesses, in both cases we are firmly in the ultrathin film regime ($d \ll \lambda_0$) where the elliptic surface PhPs hybridize creating a symmetric and antisymmetric mode.[27,50] The symmetric mode, commonly referred to as epsilon-near-zero (ENZ), is generally non-dispersive and exists at higher frequency, whereas the antisymmetric mode (Figure 2a), similar to plasmonic thin-film modes,[51] propagates with high-$k$ in thin films.[27]

The momenta of thin film PhPs depend on the thickness, the polaritonic medium permittivity, and the substrate ($\varepsilon_{sub}$) and superstrate ($\varepsilon_{sup}$) permittivities (SI Section S5).[27,30,52,53] Our experimental data (Figure 2c, squares) show an inverse relation between thickness and confinement for several flakes of HfSe$_2$ ($d$ = 47, 68, 85, and 119 nm), in agreement with simulations (Figure 2c, solid lines). The most highly confined PhPs ($\lambda_0/\lambda_p$ > 250) we could resolve in our experiments were observed for the thinnest flake of HfSe$_2$ on a Si substrate ($\varepsilon_{sub}$ ~ 12) at frequency $\omega_0$ = 162 cm$^{-1}$. There is a limit, however, to how much thinner a film can be while still supporting propagating PhPs, which is defined as traveling at least one full wavelength before its amplitude decays a factor 1/e. At comparable thicknesses to our experiments, the maximum confinement of PhPs in mid-IR materials (e.g. hBN, α-MoO$_3$) plateaus around $\lambda_0/100$.[32,33] Pushing to even thinner films does produce larger confinement,[8] but for mid-IR wavelengths this approaches the atomic limit where propagation losses are too large for observing PhP modes. It is therefore interesting to instead use larger free-space wavelengths rather than decrease the film thickness, while still maintaining a small ratio between the two (i.e. $d/\lambda_0$<<1), making the THz an ideal spectral range for testing the limits of ultraconfined PhPs. We simulate the PhP dispersion including propagation losses of a theoretical, type-II hyperbolic material and calculate the maximum confinement ($k$ at Re($k$) = Im($k$)) for three different thickness-to-wavelength ratios (Figure 2d, $d/\lambda_0$ = 2·10$^{-3}$, black; 1·10$^{-3}$, blue; and 5·10$^{-4}$, red); see SI Section S5 for details. With greater disparity between $d$ and $\lambda_0$, the achievable confinement $k/k_0$ for a propagating wave reaches higher values independent of the spectral range. The high loss regime introduced within atomically thin flakes can therefore be avoided using THz light with thicker material layers. However, we can only observe ultraconfined modes in such ultrathin films because of the exceptionally large light-matter coupling strength in HfDCs. The upper and lower bulk PhP branches, which exist at frequencies outside of the RB in contrast to surface and hyperbolic modes, are a result of the strong coupling between infrared light and the infrared-active phonons of polar dielectrics (Figure S5).[54] The Rabi splitting (2$g$) quantifies the strength of the coupling by the magnitude of the energy gap between the branches in the polaritonic dispersion, however, the difference in energy between the mid-IR and THz makes it misleading to compare the phonon excitations solely by g. It is therefore more appropriate to use a parameter that normalizes the Rabi splitting by the TO phonon frequency ($\omega_{TO}$).

$$\eta = \frac{g}{\omega_{TO}} = \frac{\sqrt{\omega_{LO}^2 - \omega_{TO}^2}}{2\omega_{TO}} \quad (1)$$

This normalized coupling strength $\eta$ relates the Rabi splitting $2g$ between the lower and upper bulk PhPs to the bare excitation frequency $\omega_{TO}$, and is commonly used in the analysis of ultrastrong coupling experiments (SI section S5).[55,56] The maximum achievable confinement is plotted as a function of $\eta$ in Figure 2d. Therein, we vary $\eta$ by decreasing $\omega_{TO}$ at constant LO phonon frequency ($\omega_{LO}$ = 100 cm$^{-1}$). At $\eta$ < 0.4, $k/k_0$ reaches modest values, but increases by a factor of 4 when $\eta$ approaches unity. It is this combination of light-matter coupling strength, hyperbolicity, and thickness-to-wavelength disparity that strongly enhances the confinement of polaritons in thin films.

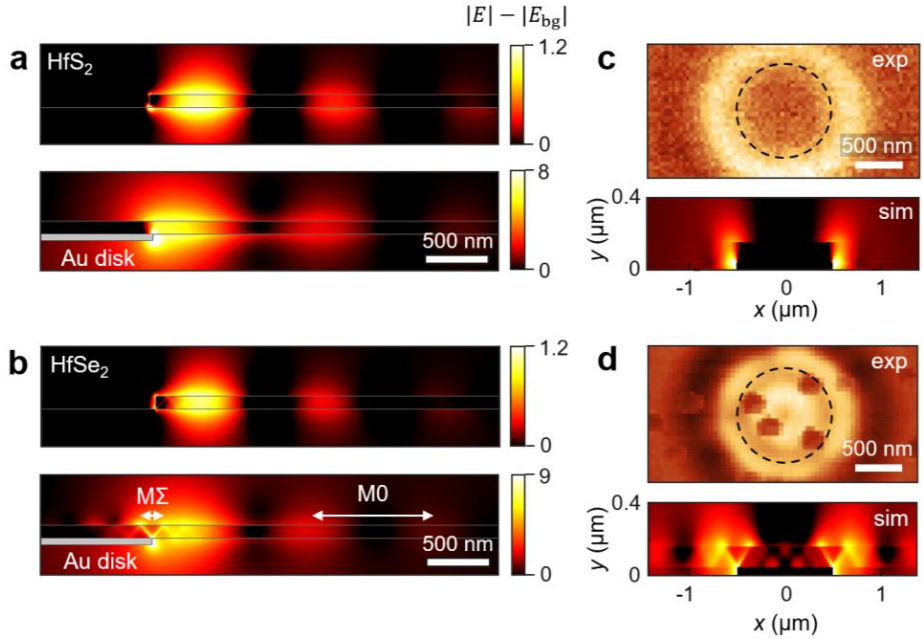

**Figure 3:** Simulated, background-subtracted electric field magnitude of phonon polaritons in thin films ($d$ = 100 nm) of (a) HfS$_2$ and (b) HfSe$_2$ on SiO$_2$ substrates. The PhP waves are excited by a p-polarized plane wave at 45° incidence angle with frequencies (a) $\omega_0$ = 240 cm$^{-1}$ and (b) $\omega_0$ = 165 cm$^{-1}$ that is scattered at the sharp edge of the thin film (top panels) or by a Au disk underneath the film (bottom panels). White double-sided arrows labeled "MΣ" and "M0" measure $\lambda_p$ of the superposition of higher-order hyperbolic modes and first-order mode in HfSe$_2$, respectively. The propagation pattern of the thin-film mode arises from the constructive and destructive interference with the plane wave. (c) Experimental top-view ($xy$) s-SNOM images (top panel, $\omega_0$ = 240 cm$^{-1}$) and simulated side-view ($xz$) of the electric field magnitude

(bottom panel, $xz$) for HfS$_2$ in a hyperlensing configuration, where a thin film is placed on top of a Au disk of radius $r = 500$ nm. (d) Similarly, s-SNOM images and simulated electric fields for HfSe$_2$ in a hyperlensing geometry ($\omega_0 = 168$ cm$^{-1}$).

Beyond the extremely compressed wavelength of polaritons observed here, hyperbolic materials can additionally support higher-order modes whose superposition leads to diffractionless, ray-like polaritons travelling at distinct frequency-dependent angles within the volume of the crystal,[16,57] see SI section S4. These effects, however, depend on the specific experimental geometry as to whether or not these polariton rays can contribute significantly to the near-field images, as illustrated in Figure 3. We illustrate the contrast in mode profiles by simulating the electric field of elliptic and hyperbolic PhPs in thin films of HfS$_2$ (Figure 3a) and HfSe$_2$ (Figure 3b), respectively. We compare edge launching for HfDC thin films placed on a SiO$_2$ substrate (Figure 3a,b top) as well as launching by a Au disk (Figure 3a,b bottom), which acts like a nanoantenna that scatters incident light much more efficiently than the thin film edge. The elliptic thin film mode in HfS$_2$ is launched, both, by the edge of the thin film and the Au disk, with identical wavelengths and similar profiles. The hyperbolic PhP in HfSe$_2$, launched by the thin film edge (Figure 3b, top), is dominated by the fundamental (M0) mode and very closely resembles its elliptic counterpart in this ultrathin film regime. However, unlike in HfS$_2$, we clearly resolve hyperbolic rays in HfSe$_2$ when placed on the Au disk, which constitute the superposition of all higher-order modes.[58] We note that a weak ray can be also seen without the Au disk substrate, which we explain to be due to direct excitation of these modes from the thin film edge (Figure 3b, top), which less efficiently scatters the incident light in comparison to the Au disk (Figure 3b, bottom) thus reducing the population of higher-order modes and only exciting the M0 mode, like those measured in Figure 1b.

Ray-like propagation of hyperbolic polaritons enables nanoscale hyperlensing and image magnification, owing to the rigid angle of polariton ray propagation.[15,16] By employing HfDCs in a hyperlensing geometry, we experimentally demonstrate THz hyperlensing in Figure 3.[15,16] Thin flakes of HfS$_2$ ($d = 115$ nm) and HfSe$_2$ (Figure 3d, $d = 115$ nm) are transferred onto a Au disk ($r = 500$ nm, $h = 40$ nm) and imaged with s-SNOM. The near-field patterns measured on the surfaces of the HfS$_2$ (Figure 3c, top, $\omega_0 = 240$ cm$^{-1}$) and HfSe$_2$ (Figure 3d, top, $\omega_0 = 168$ cm$^{-1}$) consist of bright 'hot-rings' encircling the diameter of the Au disk (Figure 3a,b top, dashed circle), consistent with similar measurements reported in hBN.[15,16] However, an additional ring appears on the HfSe$_2$ sample also inside the diameter of the Au disk, suggesting that indeed we observe THz hyperlensing for hyperbolic HfSe$_2$ (see SI section S6) for data at other frequencies. We note that the dark spots in the s-SNOM intensity of HfSe$_2$ (Figure 3d, top) emerge from oxidation defects common for transition-metal selenides,[59] but interestingly do not appear to influence the polariton propagation via scattering or other deleterious interactions.

To corroborate these experimental observations, we simulate the electric fields from a two-dimensional side-view profile (*xz*, Figure 3c,d bottom) for both cases, elliptic HfS$_2$ and hyperbolic HfSe$_2$. The Au disk scatters incident light and strongly focuses the electric field around its sharp edge, subsequently launching PhPs into the HfDC thin film above. For elliptic HfS$_2$ at that frequency, only the antisymmetric thin film mode is supported, and thus the blurred ring corresponds to the first lobe of that wave and is only observable due to the thin film nature of the layer. In contrast for hyperbolic HfSe$_2$, the gold edge couples not only to the fundamental M0 but also to higher order modes, to form hyperbolic rays[16,57] that propagate radially in both directions from the edge of the Au disk. The simulations confirm that indeed our results constitute, to the best of our knowledge, first observation of hyperlensing in the THz range in a natural crystal, here indicating the imaging of a 1000-nm diameter particle, which is 60-times smaller than the free-space wavelength (168 cm$^{-1}$).

**Outlook and Conclusion**

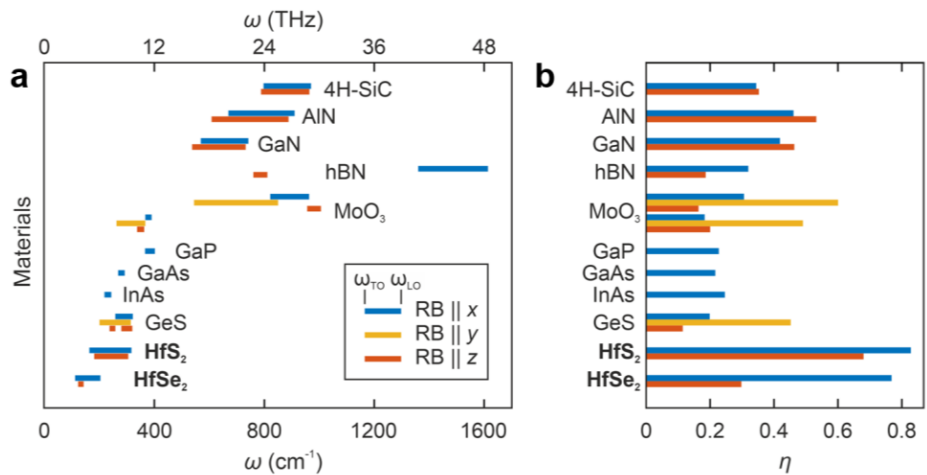

**Figure 4:** (a) Reststrahlen bands ($\omega_{TO}$ to $\omega_{LO}$) of several polar dielectric materials in the THz to mid-IR. Anisotropic materials are represented by different color bands, with isotropic materials by one (blue),

uniaxial by two (blue/red), and biaxial materials by three (blue/red/yellow) bands, respectively. (b) Normalized coupling strength $\eta$ calculated from Equation (1) for each material in (a).

Shrinking electromagnetic waves to subdiffractional mode volumes is essential for the development of nanophotonic devices, enhanced imaging capabilities, and the exploration of nanoscale phenomena. In the infrared, the mechanisms to achieve high-$k$ PhPs have primarily been reserved to identifying hyperbolic media and/or pushing light confinement to their absolute limit.[30,48] Observation of ultrahigh-$k$, broadband, elliptic and hyperbolic PhPs in $HfS_2$ and $HfSe_2$ in the THz range represents a considerable addition to the library of polaritonic materials. Moreover, it reveals the significance of the light-matter coupling strength, a material property that significantly influences the propagation of polaritons. The search for phonon polaritonic materials with ultrastrong light-matter coupling strengths has largely been focused on the mid-IR where polar dielectrics such as silicon carbide (SiC),[60] α-$MoO_3$,[36] and III-nitrides[33,61,62] display large LO-TO phonon splittings (> 100 cm$^{-1}$). The RBs of several phonon polaritonic materials are shown in Figure 4a, illustrating the difference in size between those in the mid-IR and THz. However, to fully understand the coupling strength of a polaritonic material, it is necessary to consider the magnitude of its phonon splitting relative to the TO phonon frequency. After calculating $\eta$ for the same set of dielectrics (Figure 4b), we find those values for $HfS_2$ and $HfSe_2$ to be nearly double that of the previously reported mid-IR materials, despite their smaller absolute RB widths. These large coupling strengths are the result of the exceptionally large Born effective charge magnitudes and the low TO phonon energies.[40,41]

Despite the interest in the THz community for nanophotonics such as biomolecule fingerprinting,[63] medical diagnostics,[64] and thermal management,[65] traditional optics have restricted the spatial resolution of imaging and spectroscopy due to the long free-space wavelengths. Furthermore, compressing electromagnetic fields to subdiffractional length scales results in high intensities, which introduces nonlinearities that may facilitate THz strong-field physics.[66] In this work, we imaged in real space ultrahigh-$k$ THz PhPs ($\lambda_0/\lambda_p$ > 250) in 2D vdW crystals where the thickness does not approach the atomic limit ($\geq$ 50 nm). The high normalized light-matter coupling strength of HfDCs, which facilitate large confinement while avoiding deleterious losses, are highlighted as crucial parameters to aid identification of novel polaritonic materials. We find that it is feasible to propagate THz light at the length scale of visible wavelengths, which continues to push the field of nanophotonics to its limits.

**Acknowledgements**

R.A.K acknowledges that this work was supported by a NASA Space Technology Graduate Research Opportunity. K.D.-G was supported by the Army Research Office under grant number W911NF-21-1-0119. A.S.S acknowledges support by the Department of Energy—Basic Energy Sciences under Grant DE-FG02-09ER4655. S.D. and J.D.C were supported by the Office of Naval Research MURI on Twist-Optics under grant N0001-23-1-2567. P.A.-G. acknowledges support from the European Research Council under Consolidator grant no. 101044461, TWISTOPTICS and the Spanish Ministry of Science and Innovation (State Plan for Scientific and Technical Research and Innovation grant number PID2022-141304NB-I00). MO, FGK, JW, LME, and SCK acknowledge the financial support by the Bundesministerium für Bildung und Forschung (BMBF, Federal Ministry of Education and Research, Germany, Project Grant Nos. 05K19ODB and 05K22ODA) as well as by the Deutsche Forschungsgemeinschaft (DFG, German Research Foundation) under Germany's Excellence Strategy through Würzburg-Dresden Cluster of Excellence on Complexity and Topology in Quantum Matter—ct.qmat (EXC 2147, project-id 390858490) and through the Collaborative Research Center on Chemistry of Synthetic Two-Dimensional Materials − CRC1415 (ID: 417590517). The Au disks were fabricated as part of a user project at the Center for Nanophase Materials Sciences (CNMS), which is a US Department of Energy, Office of Science User Facility at Oak Ridge National Laboratory.

**Methods**

*Sample Fabrication*

Thin flakes were fabricated by mechanical exfoliation of $HfSe_2$ and $HfS_2$ bulk crystals (HQ Graphene) using the scotch tape method.[67] Flakes were exfoliated onto either Si with a ~1μm oxide layer, which served as the $SiO_2$ substrate, or high-resistivity float zone Si.

*Hyperlens Fabrication*

Hyperlensing samples were fabricated by deterministically transferring 2D flakes of $HfS_2$ and $HfSe_2$ onto Au disks using a polymer-assisted transfer method. The disks were fabricated onto a Si substrate using electron-beam lithography and resistive Au deposition with thickness of 40 nm. HfDC flakes were mechanically exfoliated onto a Si substrate, removed using a polydimethylsiloxane (PDMS) stamp, and transferred onto the Au disks.

*FEL-SNOM*

Polariton imaging was performed using the s-SNOM at the from attocube systems AG, attached to the free-electron laser facility FELBE at the Helmholtz-Zentrum Dresden-Rossendorf, Germany.[46,47] The s-SNOM tip, with tapping frequency $\Omega \approx 160$ kHz, focuses the FEL radiation at its apex and scatters the signal into the far-field. The scattered signal is composed of multiple harmonics ($n\Omega$, $n = 1, 2, 3…$) that are demodulated from the linear far-field background. A liquid helium-cooled gallium-doped germanium photoconductive detector (QMC Instruments Ltd) was used to detect the scattered signal. For the edge-launched polaritons, the second harmonic near-field signal ($S_{O2A}$, $n = 2$) was recorded using either homodyne or self-homodyne interferometric technique, while the hyperlensing measurements were done using self-homodyne detection.[68]

*Transfer Matrix Simulations*

The dispersion maps shown in Figure 2 were calculated using a transfer matrix algorithm,[41] evaluating the imaginary component of the P-polarized reflection coefficient for evanescent wave excitation at a given in-plane momentum k. Peaks in these maps correspond to absorptive polariton resonances, [41] whereas the linewidth of these peaks in horizontal cuts (at constant frequency) relate to the propagation losses, and the linewidth in vertical cuts (at constant momentum) relate to the polariton life time. These quantities are used to estimate the maximum achievable confinement as shown in Figure 2d, see SI section S5 for details.

*Full-wave finite-element Simulations*

Full-wave finite-element simulations were conducted with Comsol Multiphysics using the electromagnetic waves, frequency domain solver of the RF module. A 2D simulation cell in the *xz* plane (vertical cross-section through the HfDC film) was constructed to simulate the hyperlensing or edge-launching experiments. Plane-wave illumination (normal incidence for hyperlensing) was used to excite phonon polaritons and the simulation cell was surrounded by perfectly matched layers. For the hyperlensing simulations in Figure 3, we used the dielectric function of HfDCs from Ref. [40] and set $\varepsilon_{\infty,z} = 7$ for $HfSe_2$. The background electric field without Au disk launcher and HfDC was subtracted.

# References

[1] A. Campion, P. Kambhampati, *Chem Soc Rev* **1998**, *27*, 241.

[2] P. Y. Chen, C. Argyropoulos, A. Alù, N. Van Hulst, *Nanophotonics* **2012**, *1*, 221.

[3] A. K. Sarychev, V. M. Shalaev, *Phys Rep* **2000**, *335*, 275.

[4] D. Yoo, D. A. Mohr, F. Vidal-Codina, A. John-Herpin, M. Jo, S. Kim, J. Matson, J. D. Caldwell, H. Jeon, N. C. Nguyen, L. Martin-Moreno, J. Peraire, H. Altug, S. H. Oh, *Nano Lett* **2018**, *18*, 1930.

[5] P. Lalanne, C. Sauvan, J. P. Hugonin, *Laser Photon Rev* **2008**, *2*, 514.

[6] A. Vakil, N. Engheta, *Science (1979)* **2011**, *332*, 1291.

[7] S. Dai, W. Fang, N. Rivera, Y. Stehle, B.-Y. Jiang, J. Shen, R. Yingjie Tay, C. J. Ciccarino, Q. Ma, D. Rodan-Legrain, P. Jarillo-Herrero, E. Hang Tong Teo, M. M. Fogler, P. Narang, J. Kong, D. N. Basov, S. Dai, J. Shen, W. Fang, J. Kong, N. Rivera, C. J. Ciccarino, P. A. Narang John Paulson, Q. Ma, D. Rodan-Legrain, P. Jarillo-Herrero, Y. Stehle, B. Jiang, M. M. Fogler, R. Y. Tay, E. H. T Teo, D. N. Basov, *Advanced Materials* **2019**, *31*, 1806603.

[8] N. Li, X. Guo, X. Yang, R. Qi, T. Qiao, Y. Li, R. Shi, Y. Li, K. Liu, Z. Xu, L. Liu, F. J. García de Abajo, Q. Dai, E. G. Wang, P. Gao, *Nat Mater* **2020**, *20*, 43.

[9] D. A. Iranzo, S. Nanot, E. J. C. Dias, I. Epstein, C. Peng, D. K. Efetov, M. B. Lundeberg, R. Parret, J. Osmond, J. Y. Hong, J. Kong, D. R. Englund, N. M. R. Peres, F. H. L. Koppens, *Science (1979)* **2018**, *360*, 291.

[10] J. Zhang, L. Zhang, W. Xu, *J Phys D Appl Phys* **2012**, *45*, 113001.

[11] T. Low, A. Chaves, J. D. Caldwell, A. Kumar, N. X. Fang, P. Avouris, T. F. Heinz, F. Guinea, L. Martin-Moreno, F. Koppens, *Nat Mater* **2017**, *16*, 182.

[12] E. Galiffi, G. Carini, X. Ni, G. Álvarez-Pérez, S. Yves, E. M. Renzi, R. Nolen, S. Wasserroth, M. Wolf, P. Alonso-Gonzalez, A. Paarmann, A. Alù, *Nat Rev Mater* **2023**, *9*, 9.

[13] D. N. Basov, M. M. Fogler, F. J. García De Abajo, *Science (1979)* **2016**, *354*, DOI 10.1126/science.aag1992.

[14] J. Rho, Z. Ye, Y. Xiong, X. Yin, Z. Liu, H. Choi, G. Bartal, X. Zhang, *Nat Commun* **2010**, *1*, 1.

[15] M. He, G. R. S. Iyer, S. Aarav, S. S. Sunku, A. J. Giles, T. G. Folland, N. Sharac, X. Sun, J. Matson, S. Liu, J. H. Edgar, J. W. Fleischer, D. N. Basov, J. D. Caldwell, *Nano Lett* **2021**, *21*, 7921.

[16]  S. Dai, Q. Ma, T. Andersen, A. S. McLeod, Z. Fei, M. K. Liu, M. Wagner, K. Watanabe, T. Taniguchi, M. Thiemens, F. Keilmann, P. Jarillo-Herrero, M. M. Fogler, D. N. Basov, *Nat Commun* **2015**, *6*, 1.

[17]  M. Autore, L. Mester, M. Goikoetxea, R. Hillenbrand, *Nano Lett* **2019**, *19*, 8066.

[18]  M. He, S. I. Halimi, T. G. Folland, S. S. Sunku, S. Liu, J. H. Edgar, D. N. Basov, S. M. Weiss, J. D. Caldwell, M. He, T. G. Folland, J. D. Caldwell, S. I. Halimi, S. M. Weiss, S. S. Sunku, S. Liu, J. H. Edgar Tim Taylor, *Advanced Materials* **2021**, *33*, 2004305.

[19]  G. Hu, J. Shen, C.-W. Qiu, A. Alù, S. Dai, G. Hu, A. Alù, C. Qiu, J. Shen, S. Dai, *Adv Opt Mater* **2020**, *8*, 1901393.

[20]  S. Dai, Q. Ma, M. K. Liu, T. Andersen, Z. Fei, M. D. Goldflam, M. Wagner, K. Watanabe, T. Taniguchi, M. Thiemens, F. Keilmann, G. C. A. M. Janssen, S. E. Zhu, P. Jarillo-Herrero, M. M. Fogler, D. N. Basov, *Nat Nanotechnol* **2015**, *10*, 682.

[21]  W. Ma, P. Alonso-González, S. Li, A. Y. Nikitin, J. Yuan, J. Martín-Sánchez, J. Taboada-Gutiérrez, I. Amenabar, P. Li, S. Vélez, C. Tollan, Z. Dai, Y. Zhang, S. Sriram, K. Kalantar-Zadeh, S. T. Lee, R. Hillenbrand, Q. Bao, *Nature* **2018**, *562*, 557.

[22]  J. D. Caldwell, L. Lindsay, V. Giannini, I. Vurgaftman, T. L. Reinecke, S. A. Maier, O. J. Glembocki, *Nanophotonics* **2015**, *4*, 44.

[23]  S. Foteinopoulou, G. C. R. Devarapu, G. S. Subramania, S. Krishna, D. Wasserman, *Phonon-Polaritonics: Enabling Powerful Capabilities for Infrared Photonics*, **2019**.

[24]  S. Adachi, *Optical Properties of Crystalline and Amorphous Semiconductors* **1999**, 33.

[25]  A. Huber, N. Ocelic, D. Kazantsev, R. Hillenbrand, *Appl Phys Lett* **2005**, *87*, DOI 10.1063/1.2032595/117557.

[26]  A. J. Huber, N. Ocelic, R. Hillenbrand, *J Microsc* **2008**, *229*, 389.

[27]  A. Mancini, L. Nan, F. J. Wendisch, R. Berté, H. Ren, E. Cortés, S. A. Maier, *ACS Photonics* **2022**, *9*, 3696.

[28]  A. M. Dubrovkin, B. Qiang, T. Salim, D. Nam, N. I. Zheludev, Q. J. Wang, *Nat Commun* **2020**, *11*, 1.

[29]  G. Álvarez-Pérez, K. V. Voronin, V. S. Volkov, P. Alonso-González, A. Y. Nikitin, *Phys Rev B* **2019**, *100*, 235408.

[30]  S. Dai, Z. Fei, Q. Ma, A. S. Rodin, M. Wagner, A. S. McLeod, M. K. Liu, W. Gannett, W. Regan, K. Watanabe, T. Taniguchi, M. Thiemens, G. Dominguez, A. H. Castro Neto, A.

Zettl, F. Keilmann, P. Jarillo-Herrero, M. M. Fogler, D. N. Basov, *Science (1979)* **2014**, *343*, 1125.

[31] T. V. A. G. de Oliveira, T. Nörenberg, G. Álvarez-Pérez, L. Wehmeier, J. Taboada-Gutiérrez, M. Obst, F. Hempel, E. J. H. Lee, J. M. Klopf, I. Errea, A. Y. Nikitin, S. C. Kehr, P. Alonso-González, L. M. Eng, *Advanced Materials* **2021**, *33*, DOI 10.1002/adma.202005777.

[32] Z. Zheng, J. Chen, Y. Wang, X. Wang, X. Chen, P. Liu, J. Xu, W. Xie, H. Chen, S. Deng, N. Xu, *Advanced Materials* **2018**, *30*, 1.

[33] J. D. Caldwell, A. V. Kretinin, Y. Chen, V. Giannini, M. M. Fogler, Y. Francescato, C. T. Ellis, J. G. Tischler, C. R. Woods, A. J. Giles, M. Hong, K. Watanabe, T. Taniguchi, S. A. Maier, K. S. Novoselov, *Nat Commun* **2014**, *5*, 1.

[34] S. Chen, P. L. Leng, A. Konečná, E. Modin, M. Gutierrez-Amigo, E. Vicentini, B. Martín-García, M. Barra-Burillo, I. Niehues, C. Maciel Escudero, X. Y. Xie, L. E. Hueso, E. Artacho, J. Aizpurua, I. Errea, M. G. Vergniory, A. Chuvilin, F. X. Xiu, R. Hillenbrand, *Nat Mater* **2023**, *8*, 1.

[35] J. Huang, L. Tao, N. Dong, H. Wang, S. Zhou, J. Wang, X. He, K. Wu, *Adv Opt Mater* **2023**, *11*, 2202048.

[36] G. Álvarez-Pérez, T. G. Folland, I. Errea, J. Taboada-Gutiérrez, J. Duan, J. Martín-Sánchez, A. I. F. Tresguerres-Mata, J. R. Matson, A. Bylinkin, M. He, W. Ma, Q. Bao, J. I. Martín, J. D. Caldwell, A. Y. Nikitin, P. Alonso-González, *Advanced Materials* **2020**, *32*, 1.

[37] T. Nörenberg, G. Álvarez-Pérez, M. Obst, L. Wehmeier, F. Hempel, J. M. Klopf, A. Y. Nikitin, S. C. Kehr, L. M. Eng, P. Alonso-González, T. V. A. G. de Oliveira, *ACS Nano* **2021**, *16*, 20174.

[38] S. Chen, A. Bylinkin, Z. Wang, M. Schnell, G. Chandan, P. Li, A. Y. Nikitin, S. Law, R. Hillenbrand, *Nat Commun* **2022**, *13*, 1.

[39] R. Xu, I. Crassee, H. A. Bechtel, Y. Zhou, A. Bercher, L. Korosec, C. W. Rischau, J. Teyssier, K. J. Crust, Y. Lee, S. N. Gilbert Corder, J. Li, J. A. Dionne, H. Y. Hwang, A. B. Kuzmenko, Y. Liu, *Nat Commun* **2024**, *15*, 1.

[40] R. A. Kowalski, J. R. Nolen, G. Varnavides, S. M. Silva, J. E. Allen, C. J. Ciccarino, D. M. Juraschek, S. Law, P. Narang, J. D. Caldwell, *Adv Opt Mater* **2022**, *10*, DOI 10.1002/adom.202200933.

[41]  S. N. Neal, S. Li, T. Birol, J. L. Musfeldt, *NPJ 2D Mater Appl* **2021**, *5*, DOI 10.1038/s41699-021-00226-z.

[42]  T. G. Folland, L. Nordin, D. Wasserman, J. D. Caldwell, *J Appl Phys* **2019**, *125*, DOI 10.1063/1.5090777.

[43]  A. Huber, N. Ocelic, T. Taubner, R. Hillenbrand, *Nano Lett* **2006**, *6*, 774.

[44]  X. Chen, D. Hu, R. Mescall, G. You, D. N. Basov, Q. Dai, M. Liu, *Advanced Materials* **2019**, *31*, 1.

[45]  S. C. Kehr, J. Döring, M. Gensch, M. Helm, L. M. Eng, *Synchrotron Radiat News* **2017**, *30*, 31.

[46]  M. Helm, S. Winnerl, A. Pashkin, J. M. Klopf, J. C. Deinert, S. Kovalev, P. Evtushenko, U. Lehnert, R. Xiang, A. Arnold, A. Wagner, S. M. Schmidt, U. Schramm, T. Cowan, P. Michel, *The European Physical Journal Plus* **2023**, *138*, 1.

[47]  J. Álvarez-Cuervo, M. Obst, S. Dixit, G. Carini, A. I. F Tresguerres-Mata, C. Lanza, E. Terán-García, G. Álvarez-Pérez, L. F. Álvarez-Tomillo, K. Diaz-Granados, R. Kowalski, A. S. Senerath, N. S. Mueller, L. Herrer, J. M. De Teresa, S. Wasserroth, J. M. Klopf, T. Beechem, M. Wolf, L. M. Eng, T. G. Folland, A. Tarazaga Martín-Luengo, J. Martín-Sánchez, S. C. Kehr, A. Y. Nikitin, J. D. Caldwell, P. Alonso-González, A. Paarmann, *Nat Commun* **2024**, *15*, 1.

[48]  A. Woessner, M. B. Lundeberg, Y. Gao, A. Principi, P. Alonso-González, M. Carrega, K. Watanabe, T. Taniguchi, G. Vignale, M. Polini, J. Hone, R. Hillenbrand, F. H. L. Koppens, *Nat Mater* **2014**, *14*, 421.

[49]  N. C. Passler, X. Ni, G. Carini, D. N. Chigrin, A. Alù, A. Paarmann, *Phys Rev B* **2023**, *107*, 235426.

[50]  N. C. Passler, C. R. Gubbin, T. G. Folland, I. Razdolski, D. S. Katzer, D. F. Storm, M. Wolf, S. De Liberato, J. D. Caldwell, A. Paarmann, *Nano Lett* **2018**, *18*, 4285.

[51]  J. J. Burke, G. I. Stegeman, T. Tamir, *Phys Rev B* **1986**, *33*, 5186.

[52]  I. H. Lee, M. He, X. Zhang, Y. Luo, S. Liu, J. H. Edgar, K. Wang, P. Avouris, T. Low, J. D. Caldwell, S. H. Oh, *Nat Commun* **2020**, *11*, 1.

[53]  S. G. Menabde, S. Boroviks, J. Ahn, J. T. Heiden, K. Watanabe, T. Taniguchi, T. Low, D. K. Hwang, N. A. Mortensen, M. S. Jang, *Sci Adv* **2022**, *8*, 627.

[54]  M. Barra-Burillo, U. Muniain, S. Catalano, M. Autore, F. Casanova, L. E. Hueso, J. Aizpurua, R. Esteban, R. Hillenbrand, *Nat Commun* **2021**, *12*, 1.

[55]  N. S. Mueller, Y. Okamura, B. G. M. Vieira, S. Juergensen, H. Lange, E. B. Barros, F. Schulz, S. Reich, *Nature* **2020**, *583*, 780.

[56]  M. Barra-Burillo, U. Muniain, S. Catalano, M. Autore, F. Casanova, L. E. Hueso, J. Aizpurua, R. Esteban, R. Hillenbrand, *Nat Commun* **2021**, *12*, 1.

[57]  H. Herzig Sheinfux, L. Orsini, M. Jung, I. Torre, M. Ceccanti, S. Marconi, R. Maniyara, D. Barcons Ruiz, A. Hötger, R. Bertini, S. Castilla, N. C. H. Hesp, E. Janzen, A. Holleitner, V. Pruneri, J. H. Edgar, G. Shvets, F. H. L. Koppens, *Nat Mater* **2024**, *23*, 499.

[58]  A. Y. Nikitin, E. Yoxall, M. Schnell, S. Vélez, I. Dolado, P. Alonso-Gonzalez, F. Casanova, L. E. Hueso, R. Hillenbrand, *ACS Photonics* **2016**, *3*, 924.

[59]  Q. Yao, L. Zhang, P. Bampoulis, H. J. W. Zandvliet, *Journal of Physical Chemistry C* **2018**, *122*, 25498.

[60]  A. Powell, J. Christiansen, R. B. Gregory, T. Wetteroth, S. R. Wilson, *Phys Rev B* **1999**, *60*, 11464.

[61]  A. T. Collins, E. C. Lightowlers, P. J. Dean, *Physical Review* **1967**, *158*, 833.

[62]  K. Torii, T. Koga, T. Sota, T. Azuhata, S. F. Chichibu, S. Nakamura, *Journal of Physics: Condensed Matter* **2000**, *12*, 7041.

[63]  J. M. Bakker, L. Mac Aleese, G. Meijer, G. von Helden, *Phys Rev Lett* **2003**, *91*, DOI 10.1103/PhysRevLett.91.203003.

[64]  R. Adato, H. Altug, *Nat Commun* **2013**, *4*, DOI 10.1038/ncomms3154.

[65]  E. Rephaeli, A. Raman, S. Fan, *Nano Lett* **2013**, *13*, 1457.

[66]  T. Kampfrath, K. Tanaka, K. A. Nelson, *Nat Photonics* **2013**, *7*, 680.

[67]  K. S. Novoselov, A. K. Geim, S. V. Morozov, D. Jiang, Y. Zhang, S. V. Dubonos, I. V. Grigorieva, A. A. Firsov, *Science (1979)* **2004**, *306*, 666.

[68]  F. Keilmann, R. Hillenbrand, *Philosophical Transactions of the Royal Society A: Mathematical, Physical and Engineering Sciences* **2004**, *362*, 787.